\documentclass[12pt]{article} 
\usepackage[xdvi]{graphicx} 
 
\begin{document}   
\newcommand{\todo}[1]{{\em \small {#1}}\marginpar{$\Longleftarrow$}}
\newcommand{\labell}[1]{\label{#1}\qquad_{#1}} 
\def\half{\mbox{\scriptsize{${{1}\over{2}}$}}}
\def\halff{\mbox{\scriptsize{${{5}\over{2}}$}}}
\def\quarter{\mbox{\scriptsize{${{1}\over{4}}$}}}
\def\eighth{\mbox{\scriptsize{${{1}\over{8}}$}}}
\def\bs{\vspace{5pt}}
\def\be{\begin{eqnarray}}
\def\ee{\end{eqnarray}}
\def\ba{\begin{array}}
\def\ea{\end{array}}
\def\bc{\begin{center}}
\def\ec{\end{center}}

\rightline{hep-th/0311271}   
\rightline{DCPT-03/57} 

\vskip 2cm 

\begin{center} 
{\Large \bf Enhan\c con Solutions: Pushing Supergravity to its Limits}
\end{center} 

\vskip 2cm   
  
\renewcommand{\thefootnote}{\fnsymbol{footnote}} \centerline{\bf
Apostolos~Dimitriadis${}^{a}$\footnote{Apostolos.Dimitriadis@durham.ac.uk},
Amanda~W.~Peet${}^{b}$\footnote{peet@physics.utoronto.ca},
Geoff~Potvin${}^{b}$\footnote{gpotvin@physics.utoronto.ca} and
Simon~F.~Ross${}^{a}$\footnote{S.F.Ross@durham.ac.uk}}
\vskip .5cm   
\centerline{ ${}^a$\it Centre for Particle Theory, Department of  
Mathematical Sciences}   
\centerline{\it University of Durham, South Road, Durham DH1 3LE, U.K.}   
\centerline{ ${}^b$ \it Department of Physics, University of Toronto,}   
\centerline{\it Toronto, Ontario M5S 1A7, Canada}   
   
\setcounter{footnote}{0}   
\renewcommand{\thefootnote}{\arabic{footnote}}   

\vskip 2cm 

\begin{abstract}   
We extend the investigation of nonextremal enhan\c cons, finding the
most general solutions with the correct symmetry and charges. There
are two families of solutions. One of these contains a solution with a
regular horizon found previously; this previous example is shown to be
the unique solution with a regular horizon. The other family
generalises a previous nonextreme extension of the enhan\c con,
producing solutions with shells which satisfy the weak energy
condition. We argue that identifying a unique solution with a shell
requires input beyond supergravity.
\end{abstract}    

\vskip 2cm

November, 2003.

\newpage  

\section{Introduction}     

The enhan\c con mechanism~\cite{JPP:enh} provides a very interesting
novel example of singularity-resolution in string
theory. Understanding the resolution of the singularity in the
original supersymmetric solution of~\cite{Behrndt:enh,Kallosh:enh}
offers us an important insight into how string theory extends the
notion of spacetime, and has been applied to obtain interesting
physical results~\cite{JM,Constable,JJ:enh1}. However, there are many
questions concerning the nature of singularities in string theory
which cannot be addressed in a supersymmetric context, so it is very
important to try to extend any singularity resolution mechanism to
address nonextremal, finite temperature
geometries. 

In~\cite{JPP:enh,JMPR:enh}, it was found that there are nonextremal
versions of the enhan\c con geometry, and it was noted that there are
two different branches of solutions: the horizon branch, which always
has a regular event horizon, and the shell branch, which always has an
enhan\c con shell outside of the horizon (if any). The horizon branch
approaches an uncharged black hole at large masses, so it is clearly
physically relevant in this regime, but no solution on this branch
exists for a finite range of masses above the BPS
solution. Furthermore, the horizon branch solution does not exhibit
the same physics as the extreme case, as it does not necessarily
involve an enhan\c con shell. The shell branch, on the other hand,
approaches the BPS solution as a parameter goes to zero, and always
involves an enhan\c con shell. It thus represents a nonextremal
generalisation of the singularity resolution in the BPS
metric. 

However, as shown in~\cite{dr}, this geometry is unphysical, as it
violates the weak energy condition (WEC). Thus, to find a nonextremal
generalisation of the enhan\c con, we must look for more general
solutions. A further motivation for looking for more general solutions
is the confusing two-branch structure in the existing solutions: near
extremality, the only solution is the shell branch, which smoothly
approaches the BPS solution of~\cite{JPP:enh}. However, far from
extremality, we would expect the horizon branch, which approaches an
uncharged black hole solution for large masses, to be the correct
solution. The transition between these two branches is an important
unresolved problem (see~\cite{es1,dr} for investigations of this
issue).

In this paper, we will extend the investigation of nonextremal
solutions in~\cite{JPP:enh,JMPR:enh}, by finding the most general
solution of the supergravity equations of motion with the correct
symmetry and charges to correspond to a nonextremal enhan\c con
solution. We will show that there are two families of asymptotically
flat solutions, corresponding to extensions of the horizon branch and
shell branch found previously. We demonstrate that the only solution
with a regular event horizon is the horizon branch solution
of~\cite{JMPR:enh}. Considering the shell branch, we show that the
general family of solutions we have constructed satisfies the WEC for
certain ranges of parameters. We then discuss the additional input
that would be required to fix these parameters to obtain a physical
solution describing a real nonextreme generalisation of the enhan\c
con.

The general solution of the supergravity equations of motion is
described in section~\ref{sec:eqs}. The physics of these solutions is
then discussed in section~\ref{sec:new}. We conclude and discuss open
issues in section~\ref{sec:concl}.

\section{Supergravity equations}
\label{sec:eqs}

Our aim is to extend previous studies of the extreme and nonextreme
enhan\c con solutions, by finding the most general solutions of the
supergravity equations consistent with the appropriate symmetries. In
this section, we will write the metric in a convenient way, and reduce
the supergravity equations of motion to a simple system of equations
for the free functions in the metric.

We want to describe a system built up from excited D-branes wrapped on
K3. As usual, we will focus mainly on the case of D6-branes, to
simplify formulae. We describe the results of the analysis for wrapped
D4- and D5-branes at the end of this section. For D6-branes, we should
consider ten-dimensional metrics which are static and have two flat
non-compact directions and a compact K3 factor along the branes. We
assume that the metric is independent of the non-compact longitudinal
directions, and that only the overall volume of the K3 varies over the
transverse space. It is then natural to proceed by Kaluza-Klein
reducing from ten to four dimensions.

In ten dimensions, we have the Type IIA 10D supergravity action (in
string frame)
\begin{equation}
\mathcal{S}_{10} = \frac{1}{2 \kappa_{10}^2}\int d^{10} x
\sqrt{-G_{10}} \left( e^{-2\Phi_{10}} \left[ R_{10} + 4 (\partial
\Phi_{10})^2 \right] - \half |F_{(2)}|^2 - \half |F_{(6)}|^2 \right)
\,.
\end{equation}
In Kaluza-Klein reducing, we write the ten-dimensional metric in an
ansatz
\begin{equation}
dS_{10}^2 = dS_4^2 + e^{B} dx_{\|}^2 + e^{D/\sqrt{2}} ds_{K3}^2,
\end{equation}
where $dx_{\|}^2 = dx_1^2 + dx_2^2$ is a flat metric on the
non-compact longitudinal directions, we assume that $F_{(6)} = f_2
\wedge \epsilon_{K3}$, where $\epsilon_{K3}$ is the volume form
determined by the unit K3 metric $ds_{K3}^2$, and we assume that $f_2$
and $F_{(2)}$ are non-zero only in the four dimensions contained in
$dS_4^2$.  Then, following the classic technique of Maharana \&
Schwarz \cite{MS}, we can obtain an action for the four-dimensional
fields,
\begin{eqnarray}
\mathcal{S}_{4} &=& \frac{1}{2\kappa_4^2}\int d^4 x \sqrt{-G_4} (
e^{-2\phi_4} \left[ R_{4} +4 (\partial \phi_4)^2 -{1 \over 2}
(\partial B)^2 - {1 \over 2} (\partial D)^2 \right] ) \\ && -\half
e^{B + \sqrt{2} D}|F_2|^2 - \half e^{B - \sqrt{2} D} |f_2|^2 ) ,
\end{eqnarray}
where the four-dimensional dilaton $\phi_4 = \Phi_{10} - B/2 -
D/\sqrt{2}$. We can convert this 4D action to Einstein frame by
writing
\begin{equation}
g_{\mu\nu} = e^{-2\phi_4} G_{\mu\nu} \,. 
\end{equation}
The result is
\begin{eqnarray}
\mathcal{S}_{4E} &=& \frac{1}{2\kappa_4^2} \int d^4 x \sqrt{-g_4}
(R_{4E} - {1 \over 2} (\partial \Phi_4)^2 -{1 \over 2} (\partial B)^2
- {1 \over 2} (\partial D)^2 \\ && - \half e^{B + \sqrt{2} D} |F_2|^2
-\half e^{B - \sqrt{2} D} |f_2|^2 ),
\end{eqnarray}
where we have defined $\Phi_4 = 2 \phi_4$ to obtain canonically
normalised kinetic terms. Henceforth, we will work in Einstein frame
for the 4D metric. 

This process of Kaluza-Klein reduction has already led to one striking
simplification: the dilaton is completely decoupled,
\begin{equation} \nabla^2 \Phi_4 = 0 \,. \end{equation}
The other two scalars have slightly more complicated behaviour:
\begin{equation}
\nabla^2 B = \half e^B [ |F_2|^2 e^{\sqrt{2}D} + |f_2|^2
e^{-\sqrt{2}D} ],
\end{equation}
\begin{equation}
\nabla^2 D = \frac{1}{\sqrt{2}} e^B [ |F_2|^2 e^{\sqrt{2}D} - |f_2|^2
e^{-\sqrt{2}D} ] \,.
\end{equation}
The equations of motion for the gauge fields take the usual form,
\begin{equation} \label{geqs}
\nabla_\mu (e^{B+\sqrt{2} D} F^{\mu\nu}) = 0, \quad \nabla_\mu
(e^{B-\sqrt{2} D} f^{\mu\nu}) = 0. 
\end{equation}
We now wish to specify our ansatz for the four-dimensional metric. We
assume that the metric is spherically symmetric in the
three-dimensional space transverse to the branes, so the metric and
scalar fields will only depend on the radial coordinate $r$ in the
transverse space. Thus, we take the metric ansatz
\begin{equation}
ds^2_{4E} = -e^{2A(r)} dt^2 + e^{2C(r)} (dr^2 + r^2 d\Omega_2^2),
\end{equation}
choosing an isotropic gauge for the radial coordinate. Since we wish
to consider a system of D6-branes, which are magnetically charged
under $F_{(2)}$, and carry an induced D2-brane charge, which is a
magnetic charge under $F_{(6)}$, we take the ansatze for the field
strengths to be
\begin{equation}
F_2 = Q_2 \epsilon_{S^2}, \quad f_2 = q_2 \epsilon_{S^2},
\end{equation}
where $\epsilon_{S^2}$ is the volume form corresponding to the unit
sphere metric $d\Omega_2^2$. As the D2-brane charge arises from a
curvature coupling of the D6-branes wrapped on K3, it is related to
the D6-brane charge through $|q_2| = (V_*/V)|Q_2|$~\cite{JPP:enh}. 
These ansatze satisfy the gauge field equations
of motion (\ref{geqs}).

With this ansatz, the Einstein equations for the four-dimensional
metric reduce to (where $'$ denotes $\partial_r$)
\begin{eqnarray} \label{eq1}
2C'' + (C')^2 + {4 \over r} C' &=& -{1 \over 4} ((\Phi_4')^2 + (B')^2
+ (D')^2) \\ && - {1 \over 4} {e^{B-2C} \over r^4} ( e^{\sqrt{2} D}
Q_2^2 + e^{-\sqrt{2} D} q_2^2), \nonumber
\end{eqnarray}
\begin{eqnarray} \label{eq2}
(C')^2 + {2 \over r} (C'+A')+2A'C' &=& {1 \over 4} ((\Phi_4')^2 +
(B')^2 + (D')^2) \\ && - {1 \over 4} {e^{B-2C} \over r^4} (
e^{\sqrt{2} D} Q_2^2 + e^{-\sqrt{2} D} q_2^2), \nonumber
\end{eqnarray}
\begin{eqnarray} \label{eq3}
A''+C'' + (A')^2 + {1 \over r} (A'+C') &=& -{1 \over 4} ((\Phi_4')^2 +
(B')^2 + (D')^2) \\ && + {1 \over 4} {e^{B-2C} \over r^4} (
e^{\sqrt{2} D} Q_2^2 + e^{-\sqrt{2} D} q_2^2), \nonumber
\end{eqnarray}
and the scalar equations become
\begin{equation} \label{eq4}
\Phi_4'' + \Phi_4' ({2 \over r} + A' + C') = 0,
\end{equation}
\begin{equation} \label{eq5}
B'' + B' ({2 \over r} + A' + C') = {1 \over 2} {e^{B-2C} \over r^4} (
e^{\sqrt{2} D} Q_2^2 + e^{-\sqrt{2} D} q_2^2),
\end{equation}
and
\begin{equation} \label{eq6}
D'' + D' ({2 \over r} + A' + C') = {1 \over \sqrt{2}} {e^{B-2C} \over
r^4} ( e^{\sqrt{2} D} Q_2^2 - e^{-\sqrt{2} D} q_2^2).
\end{equation}

We have reduced the problem of finding the general solution subject to
the assumed symmetries to solving this system of equations for the
five unknown functions $A,B,C,D,\Phi_4$. This seems like a complicated
coupled system of equations, but in fact it conceals some remarkable
simplifications. If we introduce new functions $a(r) = A+C$, $c(r) =
C+B/2$, (\ref{eq2}) + (\ref{eq3}) gives
\begin{equation} \label{apceq}
a'' + (a')^2 + {3 \over r} a' = 0,
\end{equation}
a completely decoupled equation for $a$. Similarly,
(\ref{eq1})+(\ref{eq2})+(\ref{eq5}) gives
\begin{equation} 
c'' + c' \left[{2 \over r} + a' \right] + {1 \over r}
a' =0,
\end{equation}
which can be rearranged to write 
\begin{equation} \label{cpbeq}
[ c' r^2 e^a ]' = -  r e^a a' .
\end{equation}
Similarly, (\ref{eq4}) can be rewritten as
\begin{equation} \label{phieq}
[ \Phi_4' r^2 e^a]' =0.
\end{equation}
These equations are solvable once we know $a$. Furthermore, if we
define $x_6 = -B - D/ \sqrt{2}$  and $x_2 = -B +
D/\sqrt{2} $, then $-2$(\ref{eq5}) $-\sqrt{2}$(\ref{eq6}) becomes
\begin{equation}
[ x_6' r^2 e^a]' = -{  e^{a-2c} \over r^2} Q_2^2
e^{-2x_6}.
\end{equation}
We choose to rewrite this as 
\begin{equation} \label{xeq}
r^2 e^a [ x_6' r^2 e^a]' = -  e^{2(a-c)} Q_2^2 e^{-2x_6}.
\end{equation}
Similarly, $-2$(\ref{eq5}) $+\sqrt{2}$(\ref{eq6}) can be rewritten as
\begin{equation}
r^2 e^a [ x_2' r^2 e^a]' = - e^{2(a-c)} q_2^2
e^{-2x_2}.
\end{equation}

We now have a much simplified system of equations in terms of the
functions $a,c,x_2,x_6,\Phi_4$. Before proceeding to solve these
equations, let us express our ansatz for the ten-dimensional fields in
terms of these variables for future reference:
\begin{eqnarray} \label{10dmet}
dS^2_{10} &=& - e^{\Phi_4 + 2(a-c)} e^{-{x_6 \over 2} - {x_2 \over 2}}
  dt^2 + e^{-{x_6 \over 2} - {x_2 \over 2}} dx_{\|}^2 + e^{\Phi_4 +
  2c} e^{{x_6 \over 2} + {x_2 \over 2}} (dr^2 + r^2 d\Omega_2^2) \\
  &&+ e^{{x_2 \over 2}- {x_6 \over 2}} ds^2_{K3}, \nonumber
\end{eqnarray}
with ten-dimensional dilaton
\begin{equation}
\Phi_{10} = {\Phi_4 \over 2} + {x_2 \over 4} - {3 x_6 \over 4}
\end{equation}
and gauge fields 
\begin{equation}
F_{(2)} = Q_2 \epsilon_{S^2}, \quad F_{(6)} = q_2 \epsilon_{S^2}
\wedge \epsilon_{K3}.
\end{equation}
Note the familiar way in which the functions $x_2, x_6$ appear in the
metric and dilaton.

\subsection{General solutions of the field equations}
\label{sec:solns}

We now proceed to solve the equations. The solution of (\ref{apceq})
is
\begin{equation} \label{apc}
a = \ln \left( 1 - {r_h^2 \over r^2} \right) + C_1.
\end{equation}
Then $r^2 e^a = (r^2 - r_h^2) e^{C_1}$, and we can easily see that
the solution of (\ref{phieq}) is
\begin{equation} \label{phi}
\Phi_4 = A_1 \ln \left( {r+r_h \over r-r_h} \right) + C_2,
\end{equation}
and (\ref{cpbeq}) is solved by
\begin{equation} \label{cpb}
c = 2 \ln \left( 1 + {r_h \over r} \right) + A_2 \ln \left( {r+r_h
\over r-r_h} \right) + C_3.
\end{equation}
Then
\begin{equation}
e^{2 (a-c)} = {r^4 \over (r+r_h)^4} \left( {r-r_h \over r+r_h}
\right)^{2 A_2} e^{-2C_3} \left( {r^2 - r_h^2 \over r^2} \right)^2
e^{2C_1} = \left( {r-r_h \over r+r_h} \right)^{2(A_2+1)} e^{2 (C_1
-C_3)}.
\end{equation}

Plugging this into (\ref{xeq}) gives 
\begin{equation} \label{x6eq}
(r^2 - r_h^2) \partial_r( (r^2 - r_h^2) \partial_r x_6) e^{2x_6} =
-Q_2^2 e^{-2C_3} \left( {r-r_h \over r+r_h} \right)^{2(A_2+1)},
\end{equation}  
and similarly 
\begin{equation}\label{x2eq}
(r^2 - r_h^2) \partial_r( (r^2 - r_h^2) \partial_r x_2) e^{2 x_2} =
-q_2^2 e^{-2C_3} \left( {r-r_h \over r+r_h} \right)^{2(A_2+1)}.
\end{equation}
These are non-linear equations, but nonetheless they have a
closed-form solution. To solve them, it is convenient to introduce a
new independent variable, 
\begin{equation} \label{z}
z =  \ln \left( {r - r_h \over r + r_h} \right), 
\end{equation}
so that these equations become 
\begin{equation} \label{x6eqz}
\partial_z^2 x_6 e^{2 x_6} = -{Q_2^2 e^{-2C_3} \over 4 r_h^2}
e^{2(A_2+1)z},
\end{equation}  
\begin{equation} \label{x2eqz}
\partial_z^2 x_2 e^{2 x_2} = -{q_2^2 e^{-2C_3} \over 4 r_h^2}
e^{2(A_2+1)z}.
\end{equation}
The general solutions of these equations is 
\begin{equation} \label{x6soln}
x_6 =  \ln \left( \alpha - {Q_2^2 e^{-2C_3} \over 16 r_h^2
(A_2+\gamma+1)^2 \alpha} e^{2(A_2+\gamma+1) z}\right) - \gamma z,
\end{equation}
\begin{equation} \label{x2soln}
x_2 = \ln \left( \beta - {q_2^2 e^{-2C_3} \over 16 r_h^2 (A_2+
\kappa+1)^2 \beta} e^{2(A_2+\kappa+ 1) z} \right) - \kappa z.
\end{equation}

\subsection{Other cases}

We can carry out a similar analysis for the cases of D4-branes wrapped
on K3 in IIA and D5-branes wrapped on K3 in IIB. We will just briefly
state the results, pointing out a few minor differences relative to
the D6-brane case discussed in detail above. 

For the D4-branes, we write the ten-dimensional string frame metric
in the form
\begin{eqnarray} 
dS^2_{10} &=& - e^{2a-6c} e^{-{x_4 \over 2} - {x_0 \over 2}} dt^2 +
  e^{2c} e^{{x_4 \over 2} + {x_0 \over 2}} (dr^2 + r^2 d\Omega_4^2) \\
  &&+ e^{{x_0 \over 2}- {x_4 \over 2}} ds^2_{K3}, \nonumber
\end{eqnarray}
and write the ten-dimensional dilaton as
\begin{equation}
\Phi_{10} = - {x_4 \over 4} +  {3 x_0 \over 4}
\end{equation}
and gauge fields as 
\begin{equation}
F_{(4)} = Q_4 \epsilon_{S^4}, \quad F_{(8)} = q_4 \epsilon_{S^4}
\wedge \epsilon_{K3}.
\end{equation}
We then obtain simple equations for the functions $a,c,x_4,x_0$, as in
the previous case. Note that the absence of any unwrapped directions
along the brane implies that there is one less scalar field in the
dimensional reduction here; it is the decoupled scalar that we lose. 

The general solution is 
\begin{eqnarray}
a(r) &=& \ln \left( 1-\frac{r_h^6}{r^6} \right) + C_1\,, \\ c(r) &=&
\frac{1}{3} \left[ 2 \ln \left( 1 + {r_h^3 \over r^3} \right) + A_1
\ln \left(\frac{r^3+r_h^3}{r^3-r_h^3}\right) + C_2 \right] \,, \\
x_4(r) &=& \ln \left(\alpha - \frac{Q_4^2 e^{-2C_2}}{144 r_h^6
(A_1+\gamma+1)^2 \alpha} e^{2(A_1+\gamma+1)z} \right) - \gamma z \,,
\\ x_0(r) &=& \ln \left(\beta - \frac{q_4^2 e^{-2C_2}}{144 r_h^6
(A_1+\kappa +1)^2\beta } e^{2(A_1+\kappa+1)z} \right) - \kappa z,
\end{eqnarray}
where
\begin{equation}
z = \ln \left(\frac{r^3-r_h^3}{r^3+r_h^3} \right) \,.
\end{equation}

For the case of D5-branes in type IIB, we write the ten-dimensional
string frame metric in the form
\begin{eqnarray} 
dS^2_{10} &=& - e^{2\varphi+2a-4c} e^{-{x_5 \over 2} - {x_1 \over 2}}
  dt^2 + e^{-{x_5 \over 2} - {x_1 \over 2}} dx^2+ e^{2\varphi+2c}
  e^{{x_5 \over 2} + {x_1 \over 2}} (dr^2 + r^2 
  d\Omega_3^2) \\ &&+ e^{{x_1 \over 2}- {x_5 \over 2}} ds^2_{K3},  
\nonumber
\end{eqnarray}
where $x$ is the single unwrapped brane direction, and write the
ten-dimensional dilaton as
\begin{equation}
\Phi_{10} = {3 \over 2} \varphi - {x_5 \over 2} +  { x_1 \over 2}
\end{equation}
and gauge fields as 
\begin{equation}
F_{(3)} = Q_3 \epsilon_{S^3}, \quad F_{(7)} = q_3 \epsilon_{S^3}
\wedge \epsilon_{K3}.
\end{equation}
We then obtain simple equations for the functions
$a,c,\varphi,x_5,x_1$. In this case, the combination $\varphi$
which decouples is not the same as the five-dimensional dilaton. 

The general solution is 
\begin{eqnarray}
a(r) &=& \ln \left(1-\frac{r_h^4}{r^4} \right)+C_1 \,,  \\ 
\varphi(r) &=& A_1 \ln \left(\frac{r^2+r_h^2}{r^2-r_h^2} \right)+C_2 \\ 
c(r) &=& {1 \over 2} \left[  
2 \ln \left(1+\frac{r_h^2}{r^2} \right) + A_2 \ln
\left(\frac{r^2+r_h^2}{r_2-r_h^2} \right) +C_3 \right] \,,\\ 
x_5(z) &=& \ln \left(\alpha - 
\frac{Q_3^2 e^{-2C_3-C_2}}{64 r_h^4 (A_2 + \half A_1 +\gamma+1)^2 
\alpha}
e^{2( A_2+ \half A_1+\gamma+1)z} \right) - \gamma z \,, \\ 
x_1(z) &=& \ln
\left( \beta - \frac{q_3^2 e^{-2C_3-C_2}}{64 r_h^4 (A_2 + \half A_1
+\kappa +1)^2\beta }
e^{2(A_2 + \half A_1+\kappa+1)z} \right) - \kappa z \,,
\end{eqnarray}
where 
\begin{equation}
z = \ln \left(\frac{r^2-r_h^2}{r^2+r_h^2} \right)\,.
\end{equation}
We see that the solutions obtained in both these cases are very
similar in form to the case of D6-branes.

\section{New enhan\c cons?}
\label{sec:new}

In the last section, we found the general solution of the supergravity
equations of motion subject to the symmetries associated with an
enhan\c con-like solution. The solution has a simple closed form. It
generalises the known solutions, introducing a number of constants of
integration. We would now like to see if this leads to any new
physical enhan\c con solutions.\footnote{Note that we have not
  introduced any enhan\c con shells, so at this stage we are really
  looking for more general analogues of the repulson solution---that
  is, what we are discussing is the
  solution exterior to any enhan\c con shell.} We will just discuss
the D6-brane case; the other cases will clearly be very similar.  

We first need to impose the condition of asymptotic flatness, which
will fix some of the constants. To impose asymptotic flatness, we
require that all the functions fall off as $1/r$ at large $r$. In the
case of $\Phi_4$, this corresponds to a choice of gauge, defining the
ten-dimensional dilaton so that $\Phi_{10}(\infty) = 0$. Examining
(\ref{apc},\ref{phi},\ref{cpb}), we see that this fixes $C_1 = C_2 =
C_3 = 0$. From (\ref{x6soln},\ref{x2soln}), we obtain non-trivial
equations for $\alpha$ and $\beta$,
\begin{equation}
\alpha - {Q_2^2 \over 16 r_h^2 (A_2+\gamma+1)^2 \alpha} = 1, 
\end{equation}
\begin{equation}
\beta - {q_2^2 \over 16 r_h^2 (A_2 + \kappa +1)^2 \beta} = 1, 
\end{equation}
with solutions 
\begin{equation} \label{albe}
\alpha = {1 \over 2} (1 \pm \sqrt{ 1 + {Q_2^2 \over 4
    r_h^2(A_2+\gamma+1)^2}}), \quad 
\beta = {1 \over 2} (1 \pm \sqrt{ 1 + {q_2^2 \over 4 r_h^2(A_2 +
    \kappa +1)^2}}).  
\end{equation}
It turns out to be convenient to rewrite these as 
\begin{equation} \label{alpha}
\alpha = {1 \over 4 r_h (A_2 + \gamma +1)} \left( 2 r_h (A_2 +
  \gamma +1) \pm \sqrt{ Q_2^2 + 4 r_h^2 (A_2 + \gamma +1)^2} \right), 
\end{equation}
\begin{equation} \label{beta}
\beta = {1 \over 4 r_h (A_2 + \kappa +1)} \left( 2 r_h (A_2 +
  \kappa +1)  \pm \sqrt{ q_2^2 + 4 r_h^2 (A_2 + \kappa +1)^2} \right). 
\end{equation}

Thus, the most general asymptotically flat solution is 
\begin{equation} \label{apcAF}
a = \ln \left( 1 - {r_h^2 \over r^2} \right),
\end{equation}
\begin{equation} \label{phiAF}
\Phi_4 = A_1 \ln \left( {r+r_h \over r-r_h} \right),
\end{equation}
\begin{equation} \label{cpbAF}
c = 2 \ln \left( 1 + {r_h \over r} \right) + A_2 \ln \left( {r+r_h
\over r-r_h} \right),
\end{equation}
\begin{equation} \label{x6solnAF}
x_6 =  \ln \left( \alpha - (\alpha-1) \left( {r+r_h \over r-r_h}
\right)^{-2(A_2+\gamma+1)}\right) + 
\gamma \ln \left( {r+r_h \over r-r_h} \right),
\end{equation}
\begin{equation} \label{x2solnAF}
x_2 = \ln \left( \beta - (\beta-1) \left( {r+r_h \over r-r_h}
\right)^{-2(A_2+\kappa + 1)} \right) + \kappa \ln \left( {r+r_h \over
r-r_h} \right),
\end{equation}
with $\alpha$ and $\beta$ given by (\ref{alpha},\ref{beta}). 

To begin to analyse the physics of these solutions, we note that there
are two kinds of potential singularities in the solution
(\ref{apcAF}-\ref{x2solnAF}). There is a singularity at $r=r_h$, where
$a \to -\infty$, and other functions may diverge. Since $a \to
-\infty$ gives $g_{00} \to 0$ in (\ref{10dmet}), this singularity
could correspond to an event horizon, if we choose other constants of
integration appropriately. However, there is another possible
singularity; if we choose the lower sign in either (\ref{alpha}) or
(\ref{beta}), there will be a singularity in (\ref{x6solnAF}) or
(\ref{x2solnAF}) respectively at some $r>r_h$. This type of
singularity is the analogue of the repulson singularity in the
original enhan\c con story~\cite{JPP:enh}. We see that, as in the
discussion of nonextreme enhan\c cons in~\cite{JMPR:enh}, it arises
from a discrete choice: there are different branches of
solutions. Henceforth, we will assume that we take the positive sign
in (\ref{alpha}), and we will refer to the solution where we take the
positive sign in (\ref{beta}) as the horizon branch, and to the
solution where we take the negative sign in (\ref{beta}) as the shell
branch. The shell branch solutions will only be valid outside of an
enhan\c con shell.\footnote{Solutions on the horizon branch do not
have a repulson singularity, but they may nonetheless have a
non-trivial enhan\c con shell appearing in them, if the $K3$ volume in
(\ref{10dmet}) reaches string-scale outside the horizon
(see~\cite{JMPR:enh} for details). We will ignore this issue in what
follows; similar general remarks to those we make for the nonextremal
solutions on the shell branch will apply in this case.}

\subsection{Uniqueness of the horizon branch}
\label{sec:hor}

Addressing first the horizon branch, we will see that the only
solution where the coordinate singularity at $r=r_h$ is a regular
event horizon is the horizon branch solution found previously
in~\cite{JMPR:enh}. For $r=r_h$ to be a regular horizon, we clearly
need the ten-dimensional dilaton $\Phi_{10}$ to remain finite at
$r=r_h$. We should also require that the volume of the two-sphere and
K3 components of the metric remain finite there, to avoid any
diverging curvature invariants. Furthermore, we must require that the
factor in front of the $dx_{\|}^2$ directions remain finite: as argued
in~\cite{naked}, a divergence of such a component may not lead to
diverging curvature invariants, but it does cause a divergence in
components of the curvature in a suitable orthonormal frame. Taken
together, these conditions require that $c,\Phi_4,x_2$ and $x_6$ are
finite at $r=r_h$. That is, they impose $A_1 = A_2 = \gamma = \kappa
=0$.

Thus, we have a unique solution with a regular horizon. It has 
\begin{equation}
a = \ln \left( 1 - {r_h^2 \over r^2} \right), \Phi_4 = 0, c = 2 \ln
\left( 1 + {r_h \over r} \right),
\end{equation}
\begin{equation}
x_6 = \ln \left( \alpha - (\alpha-1) \left( {r+r_h \over r-r_h}
\right)^{-2}\right) = \ln \left( {r^2 + (Q_2^2+4 r_h^2)^{1/2} r +
r_h^2 \over (r+r_h)^2} \right),
\end{equation}
\begin{equation}
x_2 = \ln \left( \beta - (\beta-1) \left( {r+r_h \over r-r_h}
\right)^{-2}\right) = \ln \left( {r^2 + (q_2^2+4 r_h^2)^{1/2} r +
r_h^2 \over (r+r_h)^2} \right), 
\end{equation}
where in the above we have used the values of $\alpha,\beta$ from
(\ref{alpha},\ref{beta}), taking the positive sign in both
equations. Using (\ref{10dmet}), this can be easily shown to be
identical to the horizon branch solution in~\cite{JMPR:enh} written in
isotropic coordinates. 

Thus, we find that the unique solution consistent with the symmetries
we expect the enhan\c con to have possessing a regular event horizon
is the horizon-branch solution found before. This is perhaps not a
surprising result, but it is quite satisfying to be able to extend the
analysis of a particular ansatz undertaken in~\cite{JMPR:enh} to a
consideration of the most general form of nonextreme enhan\c con
metric.

\subsection{Shell branch: Extremal solutions}
\label{sec:ext}

We turn now to a discussion of the shell branch. As usual in
discussions of the enhan\c con mechanism, it is useful to first
consider the extreme case, and then extend this to nonextreme
solutions. Let us therefore consider what happens to the general
solution (\ref{apcAF}-\ref{x2solnAF}) if we take $r_h=0$.

This will depend on how we take the limit. If we take $r_h \to 0$ with
$A_1, A_2, \kappa, \gamma$ held fixed, then we recover the usual
extremal solution. We will get $a = \Phi_4 = c =0$,
\begin{equation}
\alpha \approx {|Q_2| \over 4 r_h (A_2 + \gamma +1)}, \quad \beta 
	\approx {-|q_2| \over 4 r_h (A_2 + \kappa +1)}  
\end{equation}
(recalling that we are considering the shell branch, so we take the
negative sign in (\ref{beta})), which gives 
\begin{equation}
x_6 \approx \ln \left (1 +\alpha {4 (A_2 + \gamma +1) r_h \over r}
\right) \approx \ln \left( 1 + {|Q_2| \over r} \right), 
\end{equation}
\begin{equation}
x_2 \approx \ln \left (1 +\beta {4 (A_2 + \kappa +1) r_h \over r}
\right) \approx \ln \left( 1 - {|q_2| \over r} \right), 
\end{equation}
which gives us the exterior metric of the BPS enhan\c con solution
of~\cite{JPP:enh}. 

On the other hand, we could take the limit $r_h \to 0$ with
$\tilde{A}_1 = A_1 r_h$ etc held fixed, which will give a more general
extremal solution. This still has $a=0$, but now
\begin{equation} 
c = {2\tilde{A}_2 \over r},
\end{equation}
\begin{equation}
\Phi_4 = {2 \tilde{A}_1 \over r},
\end{equation}
and
\begin{equation}
x_6 =  \ln \left( \alpha - (\alpha-1) e^{-4(\tilde{A}_2+\tilde\gamma)
\over r}\right) +2 { \tilde\gamma \over r}
\end{equation}
\begin{equation} 
x_2 = \ln \left( \beta - (\beta-1) e^{-4(\tilde{A}_2+\tilde\kappa)
\over r} \right) + 2 {\tilde \kappa \over r}.
\end{equation}
In this limit, (\ref{alpha},\ref{beta}) become
\begin{equation} 
\alpha = {1 \over 4 (\tilde A_2 + \tilde \gamma)} \left( 2 (\tilde A_2
  + \tilde \gamma) + \sqrt{ Q_2^2 + 4 (\tilde A_2 + \tilde
  \gamma)^2} \right),
\end{equation}
\begin{equation} 
\beta = {1 \over 4 (\tilde A_2 + \tilde \kappa)} \left( 2 (\tilde A_2
  + \tilde \kappa) - \sqrt{ q_2^2 + 4 (\tilde A_2 + \tilde
  \kappa)^2} \right).
\end{equation}

These additional solutions look similar to the exterior solution in
the familiar BPS enhan\c con to some extent; they have a singularity
at some $r>0$, where $x_2 \to -\infty$, implying that the volume of
the K3 goes to zero. We wish to ask if we can build a physical
solution where this singularity is resolved. To resolve the
singularity, we need to be able to consistently excise the region
inside the radius where the K3 volume reaches the self-dual point with
flat space by introducing a shell of branes at this radius.

If we consider the junction between this solution and flat space, we
can define the shell stress tensor in terms of the discontinuity in the
extrinsic curvature~\cite{israel,JMPR:enh}, 
\begin{equation}
S_{AB} \equiv \frac{1}{\kappa^2} (\gamma_{AB} - G_{AB}
\gamma^C_C)
\end{equation}
where $\gamma_{AB} \equiv K^+_{AB} + K^-_{AB}$ is the jump in the
extrinsic curvature
\begin{equation}
\mathcal{K}_{AB}^{\pm} = \mp {1 \over 2} \frac{1}{\sqrt{G_{rr}}}
\frac{\partial}{\partial r } (G_{AB}) \,.
\end{equation}
Assuming the interior metric is flat, $K_{AB}^- = 0$, so $\gamma_{AB}
= K^+_{AB}$. The components of the stress tensor for a general metric
of the form (\ref{10dmet}) are then 
\begin{equation} \label{stt}
S_{tt} = {1 \over \kappa^2 \sqrt{G_{rr}}} (4c' + x_2' + x_6') G_{tt},
\end{equation}
\begin{equation}
S_{\mu\nu} = {1 \over \kappa^2 \sqrt{G_{rr}}} (2a' + 2c' + \Phi_4' +
              x_2' + x_6') G_{\mu\nu},
\end{equation}
\begin{equation}
S_{ij} = {1 \over \kappa^2 \sqrt{G_{rr}}} 2a' G_{ij},
\end{equation}
\begin{equation}
S_{ab} = {1 \over \kappa^2 \sqrt{G_{rr}}} (2a' + 2c' + \Phi_4' + x_6')
         G_{ab},
\end{equation}
where indices $\mu,\nu$ run over the non-compact longitudinal
directions, $i,j$ run over the $S^2$ directions, and $a,b$ run over
the K3 directions. We thus see that $S_{ij} = 0$ for any solution with
$r_h = 0$, as we would expect for an extremal solution. 

Since the stress tensor in the sphere directions vanishes, it is
natural to see what happens if we try to model the source for this
shell by a collection of fundamental branes, generalising the BPS
enhan\c con solution. The DBI action for wrapped D6-branes is
\begin{equation}
S = - \int_{{\cal M}_2}d^3\xi\, e^{-\Phi_{10}} (\mu_6 V(r) - \mu_2)
(-\det{G_{\mu\nu}})^{1/2}  
\label{probeaction}
\end{equation}
where ${\cal M}_2$ is the unwrapped part of the worldvolume, which
lies in six non--compact dimensions, $V(r)$ is the running volume of
the K3, and $G_{\mu\nu}$ is the induced (string frame)
metric. Plugging in the metric~(\ref{10dmet}), we obtain
\begin{equation}
S = -\int d^3 \xi e^{a-c} (\mu_6 e^{-x_6} - \mu_2 e^{-x_2}).
\end{equation}

Since the action does not couple to the 4d dilaton $\Phi_4$, it cannot
source a discontinuity in this field; thus, we must set $\tilde
A_1=0$. The action has a Lorentz symmetry relating the time direction
and the non-compact spatial directions; we can therefore only use it
as the source if the shell stress tensor also respects this symmetry,
which forces us to set $\tilde A_2 = 0$. We are then just left with
the terms coming from $x_2'$ and $x_6'$ in the stress-energy. If these
are to be sourced by the brane action, these functions need to satisfy
$x_2' e^{x_2} = $ constant, $x_6' e^{x_6} = $ constant. These
constraints force us to set $\tilde \gamma = \tilde \kappa =0$. This
gives us back the usual BPS enhan\c con solution.

Thus, while we have found additional solutions with $r_h = 0$, these
are not physical extreme enhan\c con solutions, in the sense that they
do not correspond to the geometry sourced by a collection of BPS
branes. Requiring that the shell stress tensor have the appropriate
form to correspond to the brane sources completely fixes the constants
of integration in the solution. That is, in the extreme case at least,
our usual no-hair intuition continues to hold. The additional
parameters do not actually correspond to a family of generalised
physical solutions; the only truly physical solution is the usual one.

In passing, it is interesting to note the effect of the deformations
in the more general solution on the asymptotics of the solution---in
particular, on the ADM mass. If we just consider turning the $\tilde
\kappa$ parameter on slightly, modifying the behaviour of $x_2$, its
asymptotics will be
\begin{equation}
e^{x_2} \approx \left( 1 + {4 (\beta-1) \tilde \kappa \over r} \right)
\left( 1 + {2 \tilde \kappa \over r} \right).   
\end{equation}
Assuming $\tilde \kappa \ll q_2^2$, 
\begin{equation}
\beta \approx {-|q_2| \over 4\kappa} \left( 1 - 2 {\tilde \kappa \over
  |q_2|}  \right),
\end{equation}
so 
\begin{equation}
e^{x_2} \approx 1 - {|q_2| \over r} + {4 \tilde \kappa \over r}.
\end{equation}
The effect of this will be that positive values of $\tilde \kappa$
increase the ADM mass. This teaches us two things: first, the
solutions with $\tilde \kappa \neq 0$ are clearly not supersymmetric,
since they do not saturate the BPS bound. Second, this suggests a
potentially useful way to correct the problem with the WEC in the
nonextreme case.

\subsection{Shell branch: Nonextremal solutions}
\label{sec:shell} 

Let us now consider the nonextreme shell branch, where we take $r_h
\neq 0$. We have the freedom to consider any solution in the general
family (\ref{apcAF}-\ref{x2solnAF}). However, in this section, we will
focus just on the effects of turning on the parameter $\kappa$ which
modifies the behaviour of $x_2$. The philosophy underlying this
approach is that we need to focus on a subset of the possible
deformations to keep the formulae arising in the discussion of
manageable complexity, and this seems to be the most natural
deformation to consider, since it is $x_2$ which already has `unusual'
behaviour in any shell branch solution. We will show that turning on
this deformation is sufficient to produce solutions which do not
violate the WEC.

Let us first review the argument that the WEC condition is violated in
the usual nonextremal shell branch solution~\cite{dr}. The nonextremal
solution of~\cite{JMPR:enh} is the special case of our general
asymptotically flat solution (\ref{apcAF}-\ref{x2solnAF}) where $A_1 =
A_2 = \gamma = \kappa =0$, and we take the negative sign in
(\ref{beta}). This metric then has a repulson singularity at some
$r=r_{\rm r}$, where $x_2 \to - \infty$. As in the extremal case, the
shell branch solution can apply only outside of some enhan\c con
shell, located at the radius where the volume of the K3 reaches the
self-dual point, $V = V_* = (2\pi \sqrt{\alpha'})^4$. From the metric
(\ref{10dmet}) we see that the enhan\c con radius is given by
\begin{equation} \label{enhrad}
e^{x_2 - x_6} = {V_* \over V}.
\end{equation}
For the nonextreme shell branch solution of~\cite{JMPR:enh} in our
coordinates, this becomes 
\begin{equation} \label{enhrad2}
{\left( \beta - (\beta-1) e^{2 z} \right) \over \left( \alpha
- (\alpha-1) e^{2z} \right)} = {V_* \over V}.
\end{equation}

We assume that we excise the portion of the solution inside this
radius and replace it with either flat space or a horizon branch
solution. There is then a discontinuity at this radius, corresponding
to a shell whose stress tensor is calculated as in the extremal case
in the previous subsection. Assuming the interior solution is still
flat (which maximises the shell's contribution to the overall ADM
mass), we see from (\ref{stt}) that the shell energy density is
\begin{equation}
\rho \propto - x_2' - x_6' -4c'.
\end{equation}
The $x_6'$ and $c'$ terms make positive contributions to the energy
density. However, the choice of the negative sign in (\ref{beta})
implies that $\beta<0$, and as a consequence the first term is
negative;
\begin{equation}
- x_2' = -{ 2 (1-\beta) e^{2z} \over (\beta - (\beta-1)
  e^{-2z})} \partial_r z <0.
\end{equation}
To see that this negative term dominates, we first write the first two
terms together, using (\ref{enhrad2}),
\begin{equation}
-x_2' -x_6' = -{ 2 (1-\beta) e^{2z_{\rm e}} \over (\beta - (\beta-1)
  e^{2z_{\rm e}})} \left( 1 - {(\alpha-1) \over (1 - \beta)} {V_*
  \over V} \right) \partial_r z.
\end{equation}
This expression is valid only at the enhan\c con radius $z = z_{\rm
  e}$, where (\ref{enhrad2}) is satisfied. Now
\begin{equation}
{(\alpha -1) \over  (1-\beta)} = { \sqrt{ Q_2^2 + 4r_h^2}
- 2 r_h  \over \sqrt{q_2^2 + 4 r_h^2} + 2 r_h} <
{V \over V_*},
\end{equation}
since $|Q_2|/|q_2| = V/V_*$. Thus, the first two terms together give a
negative answer. Furthermore, for this supergravity analysis to be
relevant, we need to assume that $V_* \gg V$, so that higher-order
corrections involving the K3 curvature are suppressed. This implies by
(\ref{enhrad2}) that $(\beta - (\beta-1) e^{2z_{\rm e}}) \ll 1$, so
these terms will dominate over the remaining positive term, $-4c' = {8
  r_h \over r_{\rm e}(r_{\rm e}+r_h)}$. Thus, $\rho <0$, and the shell
  violates the WEC. The usual nonextreme enhan\c con solution thus
  cannot correspond to the geometry sourced by a physical collection
  of branes. 

A primary motivation for looking for more general solutions was to see
how general this problem is. We will now show that we can produce
solutions where the shell satisfies the WEC by generalising to
non-zero values of $\kappa$. First, we
note that changing $\kappa$ will change the enhan\c con radius;
(\ref{enhrad}) now implies 
\begin{equation} \label{enhr}
{\left( \beta - (\beta-1) e^{2(\kappa+1) z} \right) \over \left( \alpha
- (\alpha-1) e^{2z} \right)} e^{-\kappa z} = {V_* \over V}.
\end{equation}
The first two terms in the energy density are then 
\begin{equation} \label{x2x6}
-x_2' - x_6' = \left[ { 2(\kappa+1) (\beta-1) e^{2(\kappa+1)z}
\over (\beta - (\beta-1) e^{2(\kappa+1)z})} + \kappa  + {
2(\alpha-1) e^{2z} \over  \left( \alpha
- (\alpha-1) e^{2z} \right)} \right] \partial_r z.
\end{equation}
Using (\ref{enhr}), we can rewrite this as
\begin{eqnarray}
-x_2' - x_6' = { -2(\kappa+1) (1-\beta) e^{2(\kappa+1)z} \over (\beta -
(\beta-1) e^{2(\kappa+1)z})} \left[ 1 - {(\alpha-1) \over
(\kappa+1)(1-\beta)} {V_* \over V} e^{-\kappa z} \right] \partial_r z
+ \kappa \partial_r z.
\end{eqnarray}
In this generalisation, it is still true that
\begin{equation}
{(\alpha -1) \over (\kappa+1) (1-\beta)} = { \sqrt{ Q_2^2 + 4r_h^2}
- 2 r_h  \over \sqrt{q_2^2 + 4(\kappa+1)^2 r_h^2} + 2(\kappa+1) r_h} <
{V \over V_*}. 
\end{equation}
However, this does not imply that the factor in square brackets in
(\ref{x2x6}) is positive. For positive $\kappa$, the factor of
$e^{-\kappa z} >1$, and it can easily be made sufficiently large to
make this factor negative, at least for small values of $r_h$.  Note
also that the additional $\kappa \partial_r z$ term is also acting in
the same direction for positive $\kappa$. Thus, the contribution of
the $x_6'$ term can dominate over that of the $x_2'$ term for suitable
values of $\kappa$, leading to a shell stress energy which satisfies
the WEC.\footnote{This seems a natural way to modify the solution to
  satisfy the WEC; however, other possibilities certainly exist. For
  example, turning on a positive $\gamma$ will modify the
  stress-energy in a very similar way, and can also lead to solutions
  which satisfy the WEC.} 

However, we still have the problem that the solution depends on
constants of integration, which seem to represent an unphysical
freedom to modify the geometry. Simply imposing the WEC cannot
completely fix the constants of integration in the solution. These
parameters are best thought of as parameterising the shell stress
tensor, and are not wholly fixed at the supergravity level, because
supergravity on its own cannot completely determine the shell stress
tensor. At the fundamental level, there should be a definite form for
this stress tensor, which will fix these parameters (possibly up to
some discrete choices). However, this will require some input from
physics beyond supergravity, which provides a real microphysical model
for the shell stress tensor, as the DBI action did in the BPS case.

Thus, we have a complete description of the solutions at the
supergravity level which satisfy the appropriate symmetry assumptions,
and we can see that some of them will satisfy the WEC, which is our
primary physics constraint on them at this level. However, since we do
not have a microphysical model for the shells in the nonextremal
cases, we cannot determine which (if any) of this family of solutions
actually correspond to physical nonextreme generalisations of the
enhan\c con mechanism.

\section{Conclusions}
\label{sec:concl}

We have been studying the extension of the enhan\c con
mechanism~\cite{JPP:enh} to nonextremal, finite temperature
geometries. In~\cite{JPP:enh,JMPR:enh}, it was found that there are
nonextremal versions of the enhan\c con geometry, and it was noted
that there are two different branches of solutions: the horizon
branch, which always has a regular event horizon, and the shell
branch, which always has an enhan\c con shell outside of the horizon
(if any).

In this paper, we have extended the work of~\cite{JMPR:enh} by finding
the most general solution consistent with the symmetries and charges
associated with the enhan\c con. These solutions represent
generalisations of the exterior geometry in the enhan\c con
solution. One of the constants of integration, $r_h$, can be
interpreted as a nonextremality parameter, so these are generally
nonextremal solutions. We find that the branch structure noted
in~\cite{JMPR:enh} arises when we impose asymptotic flatness: this
results in a quadratic equation for one of the constants of
integration, with the two roots corresponding to the horizon branch
and the shell branch.

Considering the horizon branch, and assuming that there is no shell
outside of the horizon, we showed that imposing regularity of the
solution at the event horizon fixes the remaining free parameters,
showing that the unique solution with a regular horizon is, as
expected, the horizon branch solution of~\cite{JMPR:enh}. This
solution reduces to an uncharged black hole at large mass. 

Considering the shell branch, we saw that we had a family of solutions
at $r_h=0$. On the shell branch, we are considering singular
supergravity metrics (there is a delta-function singularity at the
location of the shell), so we can no longer fix these constants of
integration by imposing regularity of the solution. However, the only
solution in this family for which the stress tensor of the shell
inferred from the supergravity solution is of the form predicted for
a collection of wrapped branes by the DBI action was the familiar BPS
solution of~\cite{JPP:enh}. Thus, we find that if we specify a
particular form for the shell stress tensor, then as expected, there
was no remaining freedom in the form of the solution; the solution is
completely described by giving its conserved charges and ADM mass.

In the nonextreme case, the shell branch solution obtained
in~\cite{JMPR:enh} is unphysical, as it violates the weak energy
condition~\cite{dr}. We have shown that this problem can be
circumvented by considering more general solutions. This provides us
with a multiparameter family of solutions which satisfy all the
constraints on physical solutions at the supergravity level.  This
freedom to add `hair' to the exterior solution arises because the form
of the shell stress tensor is not completely fixed. Indeed, the four
free parameters in the exterior solution correspond precisely to the
freedom to specify three components of the shell stress tensor and the
discontinuity in the dilaton, although the translation between the
parameters and the stress tensor is quite non-trivial. (The freedom to
specify the shell stress in the sphere directions, which is not
affected by these parameters, corresponds to the further ambiguity
previously noted in~\cite{JMPR:enh}, in the division of the energy
above extremality between the shell itself and a black hole inside the
shell.) Thus, if we had a microphysical model of the shell, we would
expect to be able to fix all of this freedom. However, this requires
further input from physics beyond supergravity, which we do not have
available for the nonextreme cases. The appearance of these parameters
thus exposes the limits of the supergravity approach to enhan\c con
physics. 

Let us reiterate the essential difference between the two branches: on
the horizon branch, we seek a smooth supergravity solution. We can
then determine the solution uniquely without requiring additional
input, as it does not involve explicit sources. On the shell branch,
the singularity can never be clothed by a horizon; we want to describe
its resolution by the expansion of the branes sourcing the
geometry. We cannot determine the appropriate geometry uniquely, as it
involves explicit sources, and we do not have a fundamental
description of those sources for the nonextremal case.\footnote{This
suggests that the most interesting case in which to investigate the
extension of singularity resolutions to near-extreme solutions will be
the mechanism of~\cite{kleb:sr}, where the singular geometry is
replaced by a smooth supergravity solution with no explicit
sources. Some investigations of this system appear
in~\cite{buchel:bh1,buchel:bh2,gubser:bh1}. It is likely that attempts
to extend other singularity resolutions that involve explicit branes,
such as~\cite{pol:sr,buchel:enh,evans:enh}, to nonextreme cases will
suffer from the difficulty we have encountered.}

In fact, our lack of understanding of the nonextremal physics goes
deeper: we cannot exclude the possibility that none of these solutions
provide an appropriate physical description of a nonextremal enhan\c
con. It is possible that the shell thickens once we add some energy to
it, invalidating the thin-shell approximation used here~\cite{dr};
alternatively, the non-abelian gauge fields which become light near
the shell may become important (this may even lead to violations of
spherical symmetry)~\cite{wijnholt}.

It is also worth noting that our study of more general solutions has
not resolved the issue of the branch structure and phase
transitions. Assuming the near-extremal behaviour is described by some
shell branch solution, while the behaviour at large masses should be
described by the horizon branch, one expects that there will be some
phase transition between the two branches as a function of
mass. Unfortunately, since we are unable to identify the correct shell
branch solution on the basis of supergravity information alone, we
cannot even set up the problem of studying this phase transition. 

\medskip
\centerline{\bf Acknowledgements}
\medskip    
    
AD is supported in part by EPSRC studentship 00800708 and by a
studentship from the University of Durham, and thanks Georgios
Tzamtzis for useful discussions. AWP is supported by NSERC of Canada,
the Canadian Institute for Advanced Research, and an Alfred P. Sloan
Foundation Research Fellowship.  GP is supported by an NSERC
Post-Graduate Scholarship and an E.C. Stevens Fellowship.  SFR is
supported by an EPSRC Advanced Fellowship.  AWP and GP would also like
to thank the Radcliffe Institute for Advanced Study for hospitality
during the early stages of this work.

\newpage
\bibliographystyle{/home/aplm/dma0sfr/tex_stuff/bibs/utphys}  
 
\bibliography{apostolos}   
    
\end{document}